\documentclass[twocolumn,aps,prc,superscriptaddress,showpacs,floatfix]{revtex4}
\usepackage{graphicx}

\begin{document}

\title{Partonic effects on higher-order anisotropic flows in relativistic
heavy-ion collisions}
\author{Lie-Wen Chen}
\thanks{On leave from Department of Physics, Shanghai Jiao Tong University,
Shanghai 200030, China}
\affiliation{Cyclotron Institute and Physics Department, Texas A\&M University, College
Station, Texas 77843-3366}
\author{C. M. Ko}
\affiliation{Cyclotron Institute and Physics Department, Texas A\&M University, College
Station, Texas 77843-3366}
\author{Zi-Wei Lin}
\affiliation{Physics Department, Ohio State University, Columbus, Ohio 43210}
\date{\today }

\begin{abstract}
Higher-order anisotropic flows $v_{4}$ and $v_{6}$ of charged hadrons in
heavy-ion collisions at the Relativistic Heavy Ion Collider are studied in a
multiphase transport model that has previously been used successfully for
describing the elliptic flow $v_{2}$ of identified hadrons in these
collisions. We find that the same parton scattering cross section of about $%
10$ \textrm{mb} used in explaining the measured $v_{2}$ of charged hadrons
can also reproduce the recent data on their $v_{4}$ and $v_{6}$ from Au + Au
collisions at $\sqrt{s}=200$ \textrm{AGeV}. It is further found that $v_{4}$
is a more sensitive probe of the initial partonic dynamics in these
collisions than $v_{2}$. Moreover, higher-order parton anisotropic flows are
non-negligible and satisfy the scaling relation $v_{n,q}(p_{T})\sim
v_{2,q}^{n/2}(p_{T})$, which leads naturally to the observed similar scaling
relation among hadron anisotropic flows when the coalescence model is used
to describe hadron production from the partonic matter.
\end{abstract}

\maketitle

\section{introduction}

There have been extensive studies on the azimuthal anisotropy of hadron
momentum distributions in the transverse plane perpendicular to the beam
direction, particularly the lower-order elliptic flow $v_{2}$, in heavy-ion
collisions at various energies \cite{reisdorf}. The hadron transverse
momentum anisotropy is generated from the pressure anisotropy in the initial
compressed matter formed in non-central heavy ion collisions \cite%
{Barrette94, Appel98} and is thus sensitive to the properties of produced
matter. This sensitivity not only exists in the larger elliptic flow \cite%
{Ollit92,Rqmd,Danie98,Zhang99,Zheng99,Voloshin03} but also in the smaller
higher-order anisotropic flows \cite{Kolb99,Teaney99,Kolb00}. Recently,
higher-order anisotropic flows $v_{4}$, $v_{6}$, and $v_{8}$ in heavy ion
collisions at the Relativistic Heavy Ion Collider (\textrm{RHIC}) were
studied in more detail using the hydrodynamical model \cite{Kolb03}, and it
was suggested that $v_{4}$ could also provide important information on the
collision dynamics during the early stage. Although the hadron anisotropic
flows become smaller with increasing orders, recent experimental results
from \textrm{STAR} Collaboration \cite{STAR03} have demonstrated that they
are still measurable.

In the present work, the anisotropic flows $v_{2}$, $v_{4}$, and $v_{6}$ of
both partons and hadrons in heavy-ion collisions at \textrm{RHIC} are
studied in a multiphase transport (\textrm{AMPT}) model that includes
initial partonic and final hadronic interactions \cite%
{Zhang:2000bd,Lin:2001cx}. Using the version with string melting, i.e.,
allowing hadrons that are expected to be formed from initial strings to
convert to their valence quarks and antiquarks \cite{Lin:2001zk,LinHBT02,ko}%
, we find that the measured data from Au + Au collisions at $\sqrt{s}=200$ 
\textrm{AGeV} can be approximately reproduced if a parton scattering cross
section of about $10$ \textrm{mb} is used in the model. Our results show
that the hadron $v_{4}$ is indeed a more sensitive probe of the initial
partonic dynamics in relativistic heavy-ion collisions than $v_{2}$. We also
find that higher-order parton anisotropic flows are non-negligible and
satisfy the scaling relation $v_{n,q}(p_{T})\sim v_{2,q}^{n/2}(p_{T})$,
which then leads to a similar scaling relation among hadron anisotropic
flows when hadron production from the partonic matter is described by the
quark coalescence model.

\section{The AMPT model with string meltings}

The \textrm{AMPT} model \cite{Zhang:2000bd,Lin:2001cx,zhang,pal} is a hybrid
model that uses minijet partons from hard processes and strings from soft
processes in the heavy ion jet interaction generator (\textrm{HIJING}) model %
\cite{Wang:1991ht} as the initial conditions for modeling heavy-ion
collisions at ultra-relativistic energies. Since the initial energy density
in Au + Au collisions at \textrm{RHIC} is much larger than the critical
energy density at which the hadronic matter to quark-gluon plasma transition
would occur \cite{Kharzeev:2001ph,zhang}, we shall use the version which
allows the melting of initial excited strings into partons \cite{Lin:2001zk}%
. In this version, hadrons, which would have been produced from the HIJING
model, are converted instead to valence quarks and/or antiquarks.
Interactions among these partons are described by the Zhang's parton cascade
(\textrm{ZPC}) parton cascade model \cite{Zhang:1997ej}. At present, this
model includes only parton-parton elastic scatterings with the in-medium
cross section 
\begin{equation}
\frac{d\sigma _{p}}{dt}=\frac{9\pi \alpha _{s}^{2}}{2}\left( 1+{\frac{{\mu
^{2}}}{s}}\right) \frac{1}{(t-\mu ^{2})^{2}},  \label{crscp}
\end{equation}%
where $\alpha _{s}=0.47$ is the strong coupling constant, and $s$ and $t$
are the usual Mandelstam variables. The effective screening mass $\mu $
depends on the temperature and density of the partonic matter but is taken
as a parameter in \textrm{ZPC} for fixing the magnitude and angular
distribution of parton scattering cross section. Since there are no
inelastic scatterings, only quarks and antiquarks from the melted strings
are present in the partonic matter. As described in Ref.\cite{Lin:2001zk},
the transition from the partonic matter to the hadronic matter is achieved
using a simple coalescence model, which combines two nearest quark and
antiquark into mesons and three nearest quarks or antiquarks into baryons or
anti-baryons that are close to the invariant mass of these partons. The
present coalescence model is thus somewhat different from the ones recently
studied extensively \cite{greco,hwa,fries}, which are used only for
producing hadrons with intermediate transverse momenta. The final-state
hadronic scatterings are then modeled by a relativistic transport (\textrm{%
ART}) model \cite{Li:1995pr}. Using parton scattering cross sections of $6$-$%
10$ \textrm{mb}, the extended \textrm{AMPT} model is able to reproduce both
the centrality and transverse momentum (below $2$ \textrm{GeV}$/c$)
dependence of the elliptic flow \cite{Lin:2001zk} and pion interferometry %
\cite{LinHBT02} measured in Au+Au collisions at $\sqrt{s}=130$ \textrm{AGeV}
at \textrm{RHIC} \cite{Ackermann:2000tr,STARhbt01}. It has also been used
for studying the kaon interferometry in these collisions \cite{lin}.

\section{anisotropic flows}

The anisotropic flows $v_{n}$ of particles are the Fourier coefficients in
the decomposition of their transverse momentum spectra in the azimuthal
angle $\phi $ with respect to the reaction plane \cite{Posk98}, i.e., 
\begin{equation}
E\frac{d^{3}N}{dp^{3}}=\frac{1}{2\pi }\frac{dN}{p_{T}dp_{T}dy}%
[1+\sum_{n=1}^{\infty }2v_{n}(p_{T},y)\cos (n\phi )]  \label{dndphi}
\end{equation}%
Because of the symmetry $\phi \leftrightarrow -\phi $ in the collision
geometry, no sine terms appear in the above expansion. For particles at
midrapidity in collisions with equal mass nuclei, which will be considered
in the present study, anisotropic flows of odd orders vanish as a result of
the symmetry $\phi \leftrightarrow \phi +\pi $. The anisotropic flows
generally depend on the particle transverse momentum and rapidity, and for a
given rapidity the anisotropic flows at transverse momentum $p_{T}$ can be
evaluated from 
\begin{equation}
v_{n}(p_{T})=\left\langle \cos (n\phi )\right\rangle ,  \label{vn1}
\end{equation}%
where $\left\langle \cdot \cdot \cdot \right\rangle $ denotes average over
the azimuthal distribution of particles with transverse momentum $p_{T}$.
The anisotropic flows $v_{n}$ can further be expressed in terms of following
single-particle averages: 
\begin{eqnarray}
v_{2}(p_{T}) &=&\left\langle \frac{p_{x}^{2}-p_{y}^{2}}{p_{T}^{2}}%
\right\rangle   \label{v2} \\
v_{4}(p_{T}) &=&\left\langle \frac{p_{x}^{4}-6p_{x}^{2}p_{y}^{2}+p_{y}^{4}}{%
p_{T}^{4}}\right\rangle   \label{v4} \\
v_{6}(p_{T}) &=&\left\langle \frac{%
p_{x}^{6}-15p_{x}^{4}p_{y}^{2}+15p_{x}^{2}p_{y}^{4}-p_{y}^{6}}{p_{T}^{6}}%
\right\rangle   \label{v6} \\
v_{8}(p_{T}) &=&\left\langle \frac{%
p_{x}^{8}-28p_{x}^{6}p_{y}^{2}+70p_{x}^{4}p_{y}^{4}-28p_{x}^{2}p_{y}^{6}+p_{y}^{8}%
}{p_{T}^{8}}\right\rangle   \label{v8}
\end{eqnarray}%
where $p_{x}$ and $p_{y}$ are, respectively, projections of particle
momentum in and perpendicular to the reaction plane.

Since the \textrm{AMPT} model also provides information on the spatial
anisotropy of colliding matter, which is responsible for generating the
momentum anisotropic flows, it is of interest to introduce the spatial
anisotropic coefficient $s_{n}$ by expressions similar to those for the
anisotropic flows $v_{n}$ but in terms of the spatial distributions of
particles in the transverse plane.

\begin{figure}[th]
\includegraphics[height=3.2in,width=3.2in]{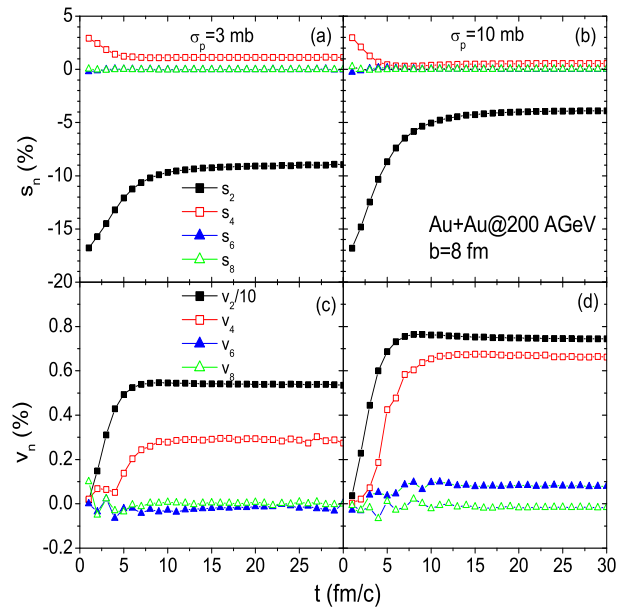}
\caption{{\protect\small (Color online) Time evolutions of spatial
anisotropic coefficients }$s_{n}${\protect\small \ and anisotropic flows }$%
v_{n}${\protect\small \ of partons in midrapidity from Au + Au collisions at 
}$\protect\sqrt{s}=200${\protect\small \ AGeV and }$b=8${\protect\small \ fm
for parton scattering cross sections }$\protect\sigma _{p}=3${\protect\small %
\ mb (left panels) and }$\protect\sigma _{p}=10${\protect\small \ mb (right
panels).}}
\label{vs2468time}
\end{figure}

Using the \textrm{AMPT} model in the string melting scenario with parton
scattering cross sections $\sigma _{p}=3$ and $10$ \textrm{mb}, we have
studied the time evolutions of $s_{n}$ and $v_{n}$ of partons in midrapidity
from Au + Au collisions at $\sqrt{s}=200$ \textrm{AGeV} and $b=8$ \textrm{fm}%
, and they are shown in Fig. \ref{vs2468time}. The spatial anisotropic
coefficients $s_{n}$ at a given time are evaluated using parton positions at
their last interactions before this time as in Ref.\cite{Lin:2001zk}. In
this way, final values of $s_{n}$ are given by the parton spatial
distributions at freeze-out. Similarly, final values of $v_{n}$ are
determined by the parton momentum distributions at freeze-out, when they are
converted to hadrons by coalescence. It is seen from Figs. \ref{vs2468time}%
(a) and (b) that the spatial elliptic deformation $s_{2}$ is initially large
and decreases with time. It reaches the saturation value at about $10$-$15$ 
\textrm{fm/}$c$ for both parton scattering cross sections. The same behavior
is observed for the fourth-order spatial anisotropic coefficient $s_{4}$
except that it reaches the saturation value earlier at about $5$ \textrm{fm/}%
$c$. The higher-order spatial anisotropic coefficients $s_{6}$ and $s_{8}$
are essentially zero. The nonzero spatial anisotropic coefficients $s_{n}$
imply that the parton spatial distribution is nonspherical at freeze-out. We
note that the larger parton scattering cross section leads to a smaller
spatial anisotropy at freeze-out.

The time evolutions of the momentum anisotropic flows $v_{n}$ of partons are
shown in Figs. \ref{vs2468time}(c) and (d) for the two parton scattering
cross sections of $3$ and $10$ \textrm{mb}, respectively. The elliptic flow $%
v_{2}$ is seen to saturate earlier in the collisions, i.e., at about $7$ 
\textrm{fm/}$c$. This is similar for both parton scattering cross sections,
except that the larger one leads to a larger elliptic flow. As to $v_{4}$ of
partons, its value reaches the final value at a later time than that for $%
v_{2}$, i.e., at about $10$ \textrm{fm/}$c$, which may be due to the fact
that the parton $v_{2}$ also affects its $v_{4}$. For other higher-order
anisotropic flows, $v_{6}$ of partons is very small, having a magnitude less
than $0.1\%$ for both parton scattering cross sections, while $v_{8}$ of
partons is essentially zero. An interesting result predicted by the \textrm{%
AMPT} model is that the parton $v_{4}$ is more sensitive to the parton cross
section than $v_{2}$. For instance, the final values of $v_{2}$ and $v_{4}$
of partons are about $5.4\%$ and $0.29\%$ , respectively, for $\sigma _{p}=3$
\textrm{mb}, and they are increased to about $7.5\%$ and $0.67\%$ ,
respectively, for $\sigma _{p}=$ $10$ \textrm{mb}. Therefore, $v_{4}$ of
partons is enhanced by a factor of about $2.3$ while their $v_{2}$ is only
enhanced by a factor of about $1.4$, when the parton cross section is
increased by a factor of about $3$.

\begin{figure}[th]
\includegraphics[height=4.2in,width=3.2in]{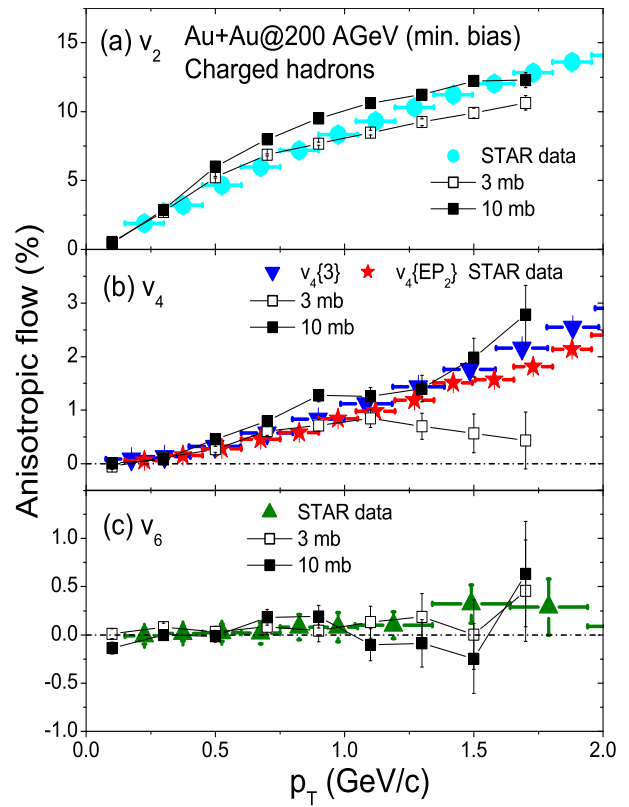}
\caption{{\protect\small (Color online) Anisotropic flows }$v_{2}$%
{\protect\small \ (a), }$v_{4}${\protect\small \ (b), and }$v_{6}$%
{\protect\small \ (c) of charged hadrons in the pseudorapidity range} $%
\left| \protect\eta \right| <1.2${\protect\small \ from minimum bias Au + Au
collisions at }$\protect\sqrt{s}=200${\protect\small \ \textrm{AGeV} as
functions of transverse momentum }$p_{T}$ {\protect\small \ for parton
scattering cross sections } $\protect\sigma _{p}=3${\protect\small \ (open
squares) and } $10${\protect\small \ (solid squares) \textrm{mb}. The
experimental data are from STAR Collaboration \protect\cite{STAR03}.}}
\label{v246ptchg}
\end{figure}

The anisotropic flows of partons are transferred to those of hadrons when
the latter are formed from the coalescence of quarks and/or antiquarks.
Although scatterings among hadrons are included in the AMPT model, they do
not affect much the hadron anisotropy flows as a result of the small spatial
anisotropy and pressure in the hadronic matter \cite{Lin:2001zk}. In Fig. %
\ref{v246ptchg}, we show the final anisotropic flows $v_{n}$ of charged
hadrons in the pseudorapidity range $\left| \eta \right| <1.2$ in minimum
bias Au + Au collisions at $\sqrt{s}=200$ \textrm{AGeV} as functions of
transverse momentum $p_{T}$ for parton scattering cross sections $\sigma
_{p}=3$ and $10$ \textrm{mb}. Also shown in Fig. \ref{v246ptchg} are recent
experimental data from \textrm{STAR} collaboration \cite{STAR03}. It is seen
that the parton scattering cross section $\sigma _{p}=3$ \textrm{mb}
underestimates the data at higher $p_{T}$ ($>1$ \textrm{GeV/}$c$) while $%
\sigma _{p}=10$ \textrm{mb} seems to give a better fit to the data. The
values of hadronic $v_{6}$ are in agreement with the data within error bars,
although they are essentially zero. Again, $v_{4}$ of charged hadrons
exhibits a stronger sensitivity to the parton cross section than their $v_{2}
$, especially at higher $p_{T}$. Therefore, $v_{4}$ of hadrons provides a
more sensitive probe of the initial partonic dynamics in relativistic
heavy-ion collisions than their $v_{2}$.

We note that the parton scattering cross section used in the present study
to reproduce the measured hadron elliptic flow is much smaller than that
needed in Ref.\cite{molnar} based only on the parton cascade model. This is
due to the fact that in Ref.\cite{molnar} the measured hadron rapidity
distribution and flow were compared to those of partons. Since the number of
partons in our model is about twice that in Ref.\cite{molnar}, we obtain a
larger effect on parton elliptic flow from parton scatterings with the same
cross section. Furthermore, the elliptic flow of hadrons that are produced
from parton coalescence is larger than that of partons if the increase of
the parton elliptic flow with transverse momentum is less than linear, which
is indeed the case as shown below in Fig.\ref{v246ptparton}.

\section{scalings in anisotropic flows}

The experimental data indicate that there is a scaling relation among hadron
anisotropic flows, i.e., $v_{n}(p_{T})\sim v_{2}^{n/2}(p_{T})$ \cite{STAR03}%
. It has been shown by Kolb \cite{kolb} that such scaling relation follows
naturally from a naive quark coalescence model \cite{Molnar03} that only
allows quarks with equal momentum to form a hadron. Denoting the meson
anisotropic flows by $v_{n,M}(p_T)$, he found that $%
v_{4,M}(p_{T})=(1/4)v_{2,M}^{2}(p_{T})$ if quarks have no higher-order
anisotropic flows. Since mesons dominate the yield of charged particles in
experimental data, the smaller scaling factor of $1/4$ than the empirical
value of about $1$ indicates that higher-order quark anisotropic flows
cannot be neglected. Including the latter contribution, one can show that 
\begin{eqnarray}
\frac{v_{4,M}(2p_{T})}{v_{2,M}^{2}(2p_{T})} &\approx &\frac{1}{4}+\frac{1}{2}%
\frac{v_{4,q}(p_{T})}{v_{2,q}^{2}(p_{T})},\allowbreak  \label{v4Mscal} \\
\frac{v_{6,M}(2p_{T})}{v_{2,M}^{3}(2p_{T})} &\approx &\allowbreak \frac{1}{4}%
\left( \frac{v_{4,q}(p_{T})}{v_{2,q}^{2}(p_{T})}+\frac{v_{6,q}(p_{T})}{%
v_{2,q}^{3}(p_{T})}\right),  \label{v6Mscal}
\end{eqnarray}%
where $v_{n,q}(p_T)$ denotes the quark anisotropic flows. The meson
anisotropic flows thus satisfy the scaling relations if the quark
anisotropic flows also satisfy such relations. 
\begin{figure}[th]
\includegraphics[height=3.0in,width=3.2in]{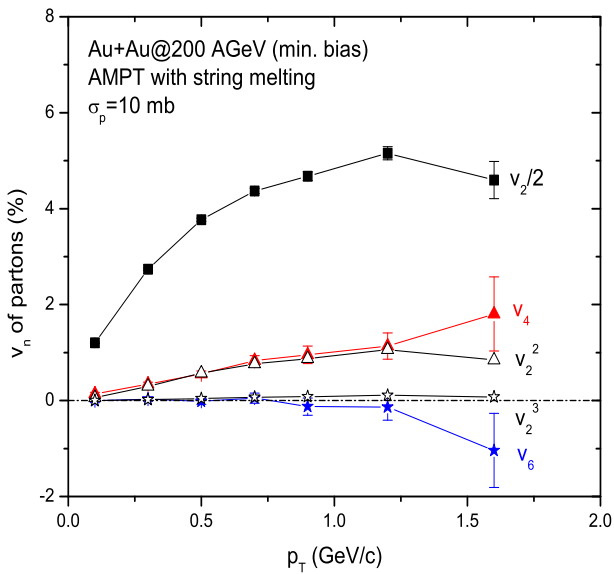}
\caption{{\protect\small (Color online) }{\protect\small \ Transverse
momentum dependence of midrapidity parton anisotropic flows }$v_{2}$%
{\protect\small , }$v_{4}${\protect\small \ and }$v_{6}${\protect\small \
from minimum bias events for Au + Au at }$\protect\sqrt{s}=200$%
{\protect\small \ \textrm{AGeV} with parton scattering cross section }$10$%
{\protect\small \ \textrm{mb}. Also plotted are }$v_{2}^{2}${\protect\small %
\ (open triangles) and }$v_{2}^{3}${\protect\small \ (open stars).}}
\label{v246ptparton}
\end{figure}

In Fig. \ref{v246ptparton}, we show the $p_{T}$ dependence of the
anisotropic flows $v_{2}$, $v_{4}$ and $v_{6}$ of midrapidity partons from
minimum bias events in Au + Au collisions at $\sqrt{s}=200$ \textrm{AGeV}
for a parton scattering cross section of $10$ \textrm{mb}. Also shown in
Fig. \ref{v246ptparton} are $v_{2}^{2}$ (open triangles) and $v_{2}^{3}$
(open stars). It is seen that the parton anisotropic flows from the \textrm{%
AMPT} model indeed satisfy the scaling relation $v_{n,q}(p_{T})\sim
v_{2,q}^{n/2}(p_{T})$. With a parton scaling factor of about $1$, we obtain
the following scaling relations among meson anisotropic flows: 
\begin{equation}
\frac{v_{4,M}(p_{T})}{v_{2,M}^{2}(p_{T})}\approx \frac{3}{4}\text{,}~~~\frac{%
v_{6,M}(p_{T})}{v_{2,M}^{3}(p_{T})}\approx \allowbreak \frac{1}{2}.
\label{v6MscalS}
\end{equation}%
The resulting hadron scaling factors of $3/4$ and $1/2$ are, however,
smaller than the one extracted from measured anisotropic flows of charged
hadrons. Since the naive quark coalescence model does not allow hadron
formation from quarks with different momenta as in more realistic quark
coalescence models \cite{alcor,greco,hwa,fries}, it is not expected to give
a quantitative description of the experimental observation. Such effects
are, nevertheless, included in the \textrm{AMPT} model, which have been
shown in Fig. \ref{v246ptchg} to reproduce the measured hadron anisotropic
flows.

\bigskip

\section{summary}

In summary, using the \textrm{AMPT} model in the string melting scenario, we
have studied the anisotropic flows $v_{2}$, $v_{4}$, and $v_{6}$ of both
partons and hadrons in heavy-ion collisions at \textrm{RHIC}. We find that
the anisotropic flows are developed early in the collision when the matter
is in the partonic stage and has appreciable spatial anisotropy. With a
parton scattering cross section of about $10$ \textrm{mb}, recent data on
hadron isotropic flows from Au + Au collisions at $\sqrt{s}=200$ \textrm{AGeV%
} can be approximately reproduced. Our results further show that $v_{4}$ is
a more sensitive probe of the initial partonic dynamics in these collisions
than $v_{2}$, and that higher-order parton anisotropic flows are
non-negligible and satisfy the scaling relation $v_{n,q}(p_{T})\sim
v_{2,q}^{n/2}(p_{T})$. Using the quark coalescence model, we then show that
the parton scaling relation leads naturally to the observed similar scaling
relation among hadron anisotropic flows. Studies of higher-order anisotropic
flows in relativistic heavy-ion collisions are thus expected to provide
useful information on the collision dynamics and the properties of the hot
dense matter during the initial stage.

\section*{Acknowledgments}

We thank Vincenzo Greco, Peter Kolb, and Ralf Rapp as well as Art Poskanzer
and Nu Xu for useful discussions. This paper was based on work supported by
the U.S. National Science Foundation under Grant No. PHY-0098805 and the
Welch Foundation under Grant No. A-1358. L.W.C. was also supported by the
National Natural Science Foundation of China under Grant No. 10105008.

\end{document}